\documentclass[manuscript, screen]{acmart}
\usepackage[utf8]{inputenc} 

\usepackage{subcaption}
\usepackage{uri}
\usepackage{todonotes}
\usepackage{booktabs}

\setcopyright{none}

\acmDOI{}

\begin{document}

\title{Data Driven Analysis of Tiny Touchscreen Performance with MicroJam}

\author{Charles P. Martin}
\orcid{0000-0001-5683-7529}
\affiliation{%
  \institution{Department of Informatics, RITMO Centre for
    Interdisciplinary Studies in Rhythm, Time and Motion, University
    of Oslo}
  \city{Oslo}
  \country{Norway}}
\email{charlepm@ifi.uio.no}

\author{Jim Torresen}
\orcid{0000-0003-0556-0288}
\affiliation{%
  \institution{Department of Informatics, RITMO Centre for
    Interdisciplinary Studies in Rhythm, Time and Motion, University
    of Oslo}
  \city{Oslo}
  \country{Norway}}
\email{jimtoer@ifi.uio.no}

\renewcommand\shortauthors{Martin, C. P. and Torresen J.}

\begin{abstract}
  The widespread adoption of mobile devices, such as smartphones and tablets, has made touchscreens a common interface for musical performance.
  New mobile musical instruments have been designed that embrace collaborative creation and that explore the affordances of mobile devices, as well as their constraints.
  While these have been investigated from design and user experience perspectives, there is little examination of the performers' musical outputs.
  In this work, we introduce a constrained touchscreen performance app, MicroJam, designed to enable collaboration between performers, and engage in a novel data-driven analysis of more than 1600 performances using the app.
  MicroJam constrains performances to five seconds, and emphasises frequent and casual music making through a social media-inspired interface.
  Performers collaborate by replying to performances, adding new musical layers that are played back at the same time.
  Our analysis shows that users tend to focus on the centre and diagonals of the touchscreen area, and tend to swirl or swipe rather than tap.
  We also observe that while long swipes dominate the visual appearance of performances, the majority of interactions are short with limited expressive possibilities. 
  Our findings are summarised into a set of design recommendations for MicroJam and other touchscreen apps for social musical interaction.
\end{abstract}



\maketitle


\section{Introduction}\label{sec:introduction}

Popular social media apps for mobile devices have allowed millions of
users to engage with creative production of images and text. These
devices' cameras, touch-screens, powerful processors, and portability
suggest on-the-go creativity, and it would appear that straightforward
sharing with friends, or a wider network of followers, is a key factor
in encouraging users to create content of all forms. Given the many
affordances of mobile devices, it has been well noted that they are
suitable platforms for mobile music making~\cite{Tanaka:2010sp}.
Despite many creative mobile digital musical instruments (DMIs)
appearing in recent years, we have yet to see the widespread adoption
of musical creation as an integrated element of social media.
Furthermore, few musical apps have attempted to emphasise ensemble,
rather than individual, performance, even though group music-making is
often seen as a valuable social activity.

\begin{figure}
  \centering
    \includegraphics[width=1.0\columnwidth]{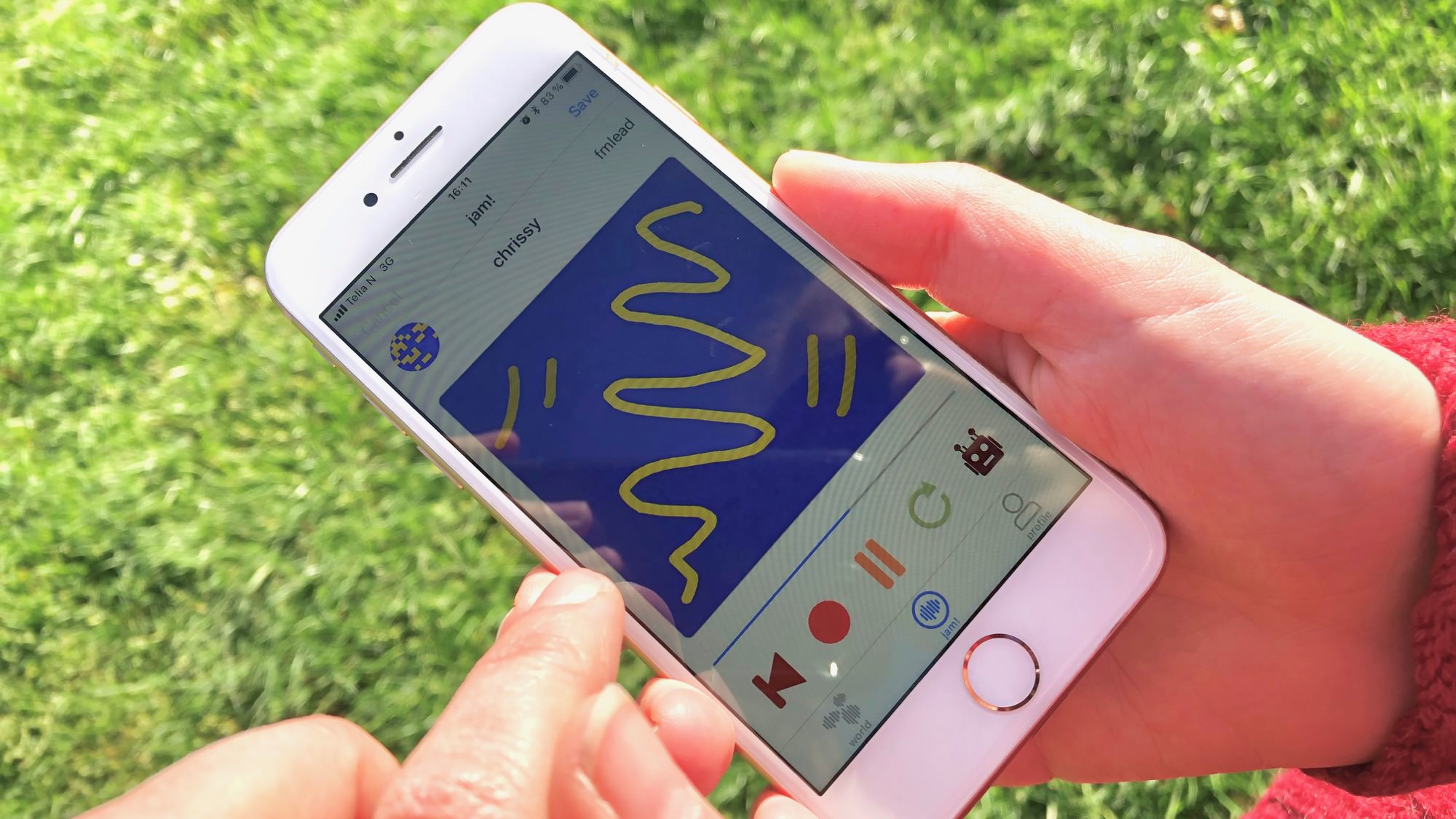}
    \caption{MicroJam allows users to create short touch-screen
      performances that are shared to other users. Replies to
      performances form distributed and asynchronous duets. A video
      demo of MicroJam is available online:
      \url{https://doi.org/10.5281/zenodo.797119}} 
  \label{fig:microjam-action}
\end{figure}

In this article, we present the design for \emph{MicroJam}\footnote{Source code and further information about MicroJam is
  available online~\cite{Martin:2017aa}.}, a
collaborative and social mobile music-making app, and an analysis of
more than 1600 touchscreen performances that have been created so
far. The design of
MicroJam emphasises casual, frequent, and social performance. As shown
in Figure \ref{fig:microjam-action}, the app features a very simple
touch-screen interface for making electronic music where skill is not
a necessary prerequisite for interaction. MicroJam departs from other
touchscreen instruments by imposing limits on the musical compositions
that are possible; most importantly, performances are limited to five
seconds in length. These ``tiny'' performances are uploaded
automatically to encourage improvisation and creation rather than
editing. Users can reply to others' performances by recording a new
layer, this combines social interaction with ensemble music making.

In Section \ref{background} we will motivate MicroJam's design with a
discussion of music-making in social media, and the possibilities for
asynchronous and distributed collaborations with mobile musical
interfaces. In Section~\ref{microjam} we will describe the app's
design and the interactive music mappings of the eight synthesised
instruments that are made available in its interface. We also
formalise the concept of ``tiny'' touchscreen performances. In
Section~\ref{studying-tiny-performances} we examine a dataset of more
than 1600 tiny performances saved on the app's cloud database by
testers and early users. We consider this dataset from several levels
of abstraction: individual touchscreen interaction events, aggregated
touchscreen gestures, and whole performances. This analysis allows us
to draw conclusions about how these users perform in MicroJam, and to
characterise the musical behaviour in the tiny performance format.
While our app design draws on themes introduced by other authors,
this is the first time that a systematic and data-driven analysis of a
large dataset of touchscreen performances has been published for such
a system. Our findings, then, will be useful for revisions of
MicroJam, and could also inform the design of other touchscreen music
applications, as well as studies of other types of DMIs.

This article is a revised and extended version of a previous
conference paper ``Exploring Social Mobile Music with Tiny
Touch-Screen Performances''~\cite{Martin:2017ac}. That paper
introduced the prototype design of MicroJam and motivated the idea of
social music making. In this new research, we present the fully
developed MicroJam app, and empirical research based on examination of
more than 1600 touchscreen performances made using the app.

\section{Background}\label{background}

Commodity mobile devices such as smartphones and tablets have often
been reframed as DMIs for research, artistic exploration, and
entertainment, forming the field of mobile music~\cite{Gaye:2006qy}.
The sensors and multitouch screens of smartphones provide many
affordances for new kinds of musical
software~\cite{essl2009interactivity, Tanaka:2010sp} and the ubiquity
of these devices increases the possibility of musical participation by
a wide audience~\cite{Essl:2018aa}. Most mobile music DMIs have used
the touchscreen as an expressive musical controller. Some DMIs, such as
Magic Fiddle~\cite{Wang:2011kx} imitate existing musical
instrument, but others, such as Crackle~\cite{2011-Crackle-Reus}, or
TC-11~\cite{Schlei:2012zr}, have defined new ways to connect
interaction on the touchscreen to sound synthesis algorithms.

While the design and evaluation of mobile music DMIs have been
reported in the academic literature (see, e.g., \cite{Essl:2018aa,
  John:2013aa, Gaye:2006qy} for surveys), few analyses exist of the
\emph{music} created with these systems. Evaluations of these systems
tend to focus on either the design and gesture-to-sound mappings
(e.g., \cite{dAlessandro:2012}), or on the experience of the musicians using the apps
(e.g., \cite{martin:2016vn}). It has previously been argued that archives of
touchscreen control data can go beyond audio in terms of analysis of
tablet musical performances~\cite{Martin:2016rm}. Similar data from
motion capture systems has previously been used as a basis for
analysis of human interactions between movement and
sound~\cite{Kelkar:2018aa}. In this work, we perform an analysis on
mobile app touchscreen data to understand how users interact musically
with MicroJam.

\subsection{Social Music-Making and Constraints}\label{social}

Many social media platforms emphasise the value of constrained
contributions by users. Twitter famously limited written notes to 140
characters~\cite{Gligoric:2018aa}, while Instagram constrained images
to square format, and the (now defunct) Vine platform only allowed six
second micro-videos~\cite{Redi:2014aa}. Constraints such as these are
often thought to lead to increased variability and creativity in the
arts~\cite{Stokes:2008aa} as well as in DMIs~\cite{Gurevich:2012xy}.  
Posts in these services are intended to be
frequent, casual, and ephemeral, and it could be that these constrained
formats have helped these apps to attract millions of users and
encourage their creativity in the written word or photography. While
social media is often used to promote music~\cite{Dewan:2014aa}, music
\emph{making} has yet to become an important creative part of the
social media landscape.

While music is often seen as an activity where accomplishment takes
practice and concerted effort, casual musical experiences are
well-known to be valuable and rewarding creative activities.
Accessible music making, such as percussion drum circles, can be used
for music therapy~\cite{Scheffel:2014aa}. DMIs such as augmented
reality instruments~\cite{Correa:2009aa} and touch-screen
instruments~\cite{Favilla:2013aa} have also been used for therapeutic
and casual music-making. In the case of DMIs, the interface can be
designed to support creativity of those without musical experience or
with limitations on motor control. Apps such as
\emph{Ocarina}~\cite{Wang:2014ul} and \emph{Pyxis
  Minor}~\cite{Barraclough:2015aa} have shown that simple touch-screen
interfaces can be successful for exploration by novice users as well
as supporting sophisticated expressions by practised performers.

Some mobile music apps have included aspects of social music-making.
Smule's \emph{Leaf Trombone} app introduced the idea of a ``world
stage''~\cite{Wang:2015ab}. In this app, users would perform
renditions of well-known tunes on a unique trombone-like DMI. Users
from around the world were then invited to critique renditions with
emoticons and short text comments. World stage emphasised the idea
that while the accuracy of a rendition could be rated by the
computer, only a human critic could tell if it was ironic or funny.
Indeed, \emph{Leaf Trombone}, and other Smule apps have made much
progress in integrating musical creation with social media.

\subsection{Jamming through Space and Time}\label{spacetime}

\begin{table}[t]
\centering
\caption{Ensemble performances typically occur with all participants
  in the same time and place; however, other situations are possible.
  MicroJam focuses on performances that are distributed (different
  place) and asynchronous (different time).}
\label{tab:time-space-matrix}
\begin{tabular}{|p{4cm}|p{4cm}|p{4cm}|}
\hline
  \textbf{Same Location}      & \emph{Mobile Ensembles}                      & \emph{Locative Performance}\\
                              & Mobile Phone Orchestra~\cite{oh2010evolving} & AuRal~\cite{Allison:2012} \\
                              & Ensemble Metatone~\cite{Martin:2015jk}       & Tactical Sound Garden Toolkit~\cite{Shepard:2007aa}\\
  \hline
  \textbf{Different Location} & \emph{Networked Performance}                 & Glee Karaoke~\cite{Hamilton:2011aa} \\
                              &                                              & Mobile DAWs \\
                              & Magic Piano~\cite{Wang:2016aa}               & \textbf{MicroJam Performances}. \\
  \hline
                              & \textbf{Same Time}                           & \textbf{Different Time}\\
  \hline
\end{tabular}
\end{table}

While performance and criticism is an important social part of
music-making, true musical collaboration involves performing music
together. These experiences of group creativity can lead to the
emergence of qualities, ideas, and experiences that cannot be easily
explained by the actions of the individual
participants~\cite{Sawyer:2006qq}. Mobile devices have often been used
in ensemble situations such as \emph{MoPho} (Stanford Mobile Phone
Orchestra)~\cite{oh2010evolving},
\emph{Viscotheque}~\cite{Swift:2013xy}, \emph{Pocket
  Gamelan}~\cite{Greg-Schiemer:2007gf},
\emph{ChoirMob}~\cite{dAlessandro:2012} and \emph{Ensemble
  Metatone}~\cite{Martin:2015jk}; however, in these examples, the
musicians played together in a standard concert situation.

Given that mobile devices are often carried by users at all times, it
would be natural to ask whether mobile device ensemble experiences can
be achieved even when performers are not in a rehearsal space or
concert venue. Could users contribute to ensemble experiences at a
time and place that is convenient to them? The use of computer
interfaces to work collaboratively even when not in the same space and
time has been extensively discussed. In HCI, groupware systems have
been framed using a time-space matrix to address how they allow users
to collaborate in the same and different times and
places~\cite{Greenberg:1998yq}. For many work tasks, it is now common
to collaborate remotely and at different times using tools such as
Google Docs or Git; however, distributed and asynchronous musical
collaboration is not as widely accepted.

In Table \ref{tab:time-space-matrix}, we have applied the time-space
matrix to mobile musical performance. Conventional collaborative
performances happen at the same time and location. Even with mobile
devices, most collaboration has occurred in this configuration.
Collaborations with performers distributed in different locations but
performing at the same time are often called networked musical
performances~\cite{Carot:2007aa}. Early versions of Smule's \emph{Magic
  Piano}~\cite{Wang:2016aa} iPad app included the possibility of
randomly assigned, real-time duets with other users. Networked
performances are also possible with conventional mobile DMIs and
systems for real-time audio and video streaming.

Performance with participants in different times, the right side of
Table \ref{tab:time-space-matrix}, are less well-explored than those
on the left. One stream of mobile music, locative
performance~\cite{Behrendt:2012aa}, has led to works that emphasise
geographical location as an important input to a musical process such
as \emph{Location33}~\cite{Carter2005},
\emph{Net\_d\'erive}~\cite{Tanaka:2008la}, or \emph{Sonic
  City}~\cite{GayeHolmquistMaze-SonicCity}. In other works, such as
\emph{AuRal}~\cite{Allison:2012}, or \emph{Tactical Sound
  Garden}~\cite{Shepard:2007aa}, users' interactions were stored at
their location, allowing collaborations in a certain space, but
separate in time.

The final area of the matrix involves music-making with performers
that are in different places and different times. \emph{Glee
  Karaoke}~\cite{Hamilton:2011aa} allows users to upload their sung
renditions of popular songs, and add layers to other performers'
contributions. The focus in this app, however, is on singing along
with a backing track, and the mobile device does not really function
as a DMI but an audio recorder. These limitations rule out many
musical possibilities such as assembling orchestras of many remote
participants and improvisation. More conventional DAWs (digital audio
workstations) are also available on mobile
devices~\cite{Meikle:2016aa}. Some of these apps (e.g., KORG's
\emph{Gadget}) offer social or collaboration features, such as
uploading whole tracks or short audio clips that other users can
incorporate into their compositions. Our app, MicroJam, fits into this
lower-right quadrant, and is distinguished from these other examples
due to its focus on constrained and ephemeral music making, as well as
online collaboration. In the next section, we will describe how this
new app enables distributed and asynchronous collaboration on original
musical material.

\section{MicroJam}\label{microjam}

MicroJam is an app for creating, sharing, and collaborating with tiny
touch-screen musical performances. This app has been specifically
created to interrogate the possibilities for collaborative mobile
performance that spans space and time. While these goals are lofty,
the design of MicroJam has been kept deliberately simple. The main
screen (see Figure \ref{fig:microjam-lineup}) recalls social-media
apps for sharing images. Musical performances in MicroJam are limited
to very short interactions, encouraging frequent and ephemeral
creative contribution. MicroJam is an iOS app written in Swift and
uses Apple's CloudKit service for backend cloud storage. The source
code is freely available for use and modification by other researchers
and performers~\cite{Martin:2017aa}. In this section we will discuss
the design of the app, the format of the tiny musical performances
that can be created, and the synthesised instruments that are
available.

\subsection{Design}

\begin{figure}
  \centering
  \includegraphics[width=1\textwidth]{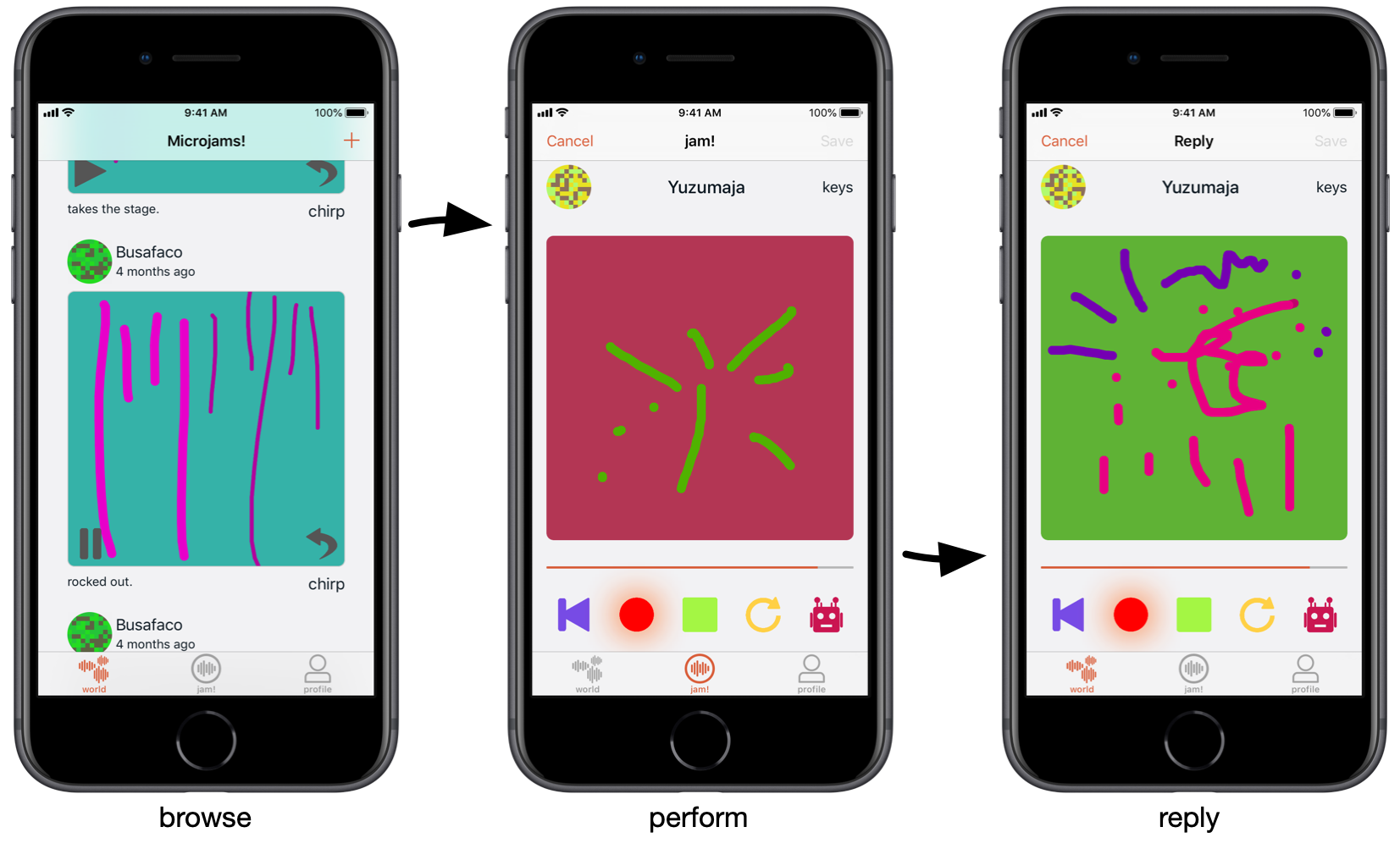}
  \caption{The MicroJam interface allows users to browse a timeline of
  performances (left), create new performances (centre), and reply, or
play in duet, with previously recorded performances (right).}
  \label{fig:microjam-lineup}
\end{figure}

MicroJam allows users to do three primary activities (shown in Figure
\ref{fig:microjam-lineup}): browse and listen to other users'
performances, create and share new performances using the touch
screen; and record layers on top of previously shared performances.
The interface for creative new performances is in the centre of Figure
\ref{fig:microjam-lineup} and is called \emph{jam!}. This screen
features a square touch performance area which is initially blank.
Tapping, swirling, or swiping anywhere in this area will create sounds
and also start recording touch activity. All touch interaction in this
area is visualised with a simple paint-style drawing that follows the
user's touches. Touch interactions are simultaneously sent to a
synthesised instrument to be sonified. After five seconds of touch
interaction, the recording is automatically stopped (although the
performer can continue to interact with the touch area). The recording
can be subsequently replayed by tapping a ``play'' icon, or
looped with the circular arrow icon. Users of MicroJam can choose the
instrument used to sonify their interactions in the jam interface from
a button in the top right. These synthesised instruments map a stream
of touchscreen events---the location of a touch, and whether it is the
start (touch down), or continuation (touch moved) of a previous
gesture---to sound. The timbres and synthesis mappings are different
for each instrument and described in Section \ref{instruments} below.

Previously recorded performances, and those recorded by other users
and downloaded from the server, are listed in the \emph{world} screen
as shown on the left side of Figure \ref{fig:microjam-lineup}. Each
performance is represented by a visual trace of the touch-drawing
captured during recording and the contributor's
online handle. Any one of these performances can be played back
in place in the world screen. When playing back, both the sound and
visualised touch-interactions are replayed in the touch-area.

When viewing a previously saved performance in the world screen, the
user can tap the reply icon (a curved arrow), to open a new layer on
top of the recording. As shown in the right side of Figure
\ref{fig:microjam-lineup}, the previous as well as current
touch-visualisations are shown and each layer is sonified separately.
Multiple replies in MicroJam are possible which can result in several
layers of performances being played back at once, allowing complex
compositions.


\subsection{Tiny touchscreen performances}\label{tiny-performance}

MicroJam is intended to provide a musical experience where constraints
are applied to the user's interaction to increase their creativity and
lower the barriers of entry for musical performance. We argued in
Section \ref{social} that constraints in social creativity systems
could actually enhance users' creative power. In the context of a
musical app, these constraints could lead to more frequent
interactions and possibly higher creativity due to the lower stakes
and effort. Musical interactions in MicroJam are similarly constrained
to be \emph{tiny} touch-screen performances as they are limited in the
area and duration of interaction. We define a tiny performance
as follows:

\begin{enumerate}
  \item All touches take place in a square subset of the touch-screen.
  \item Duration of the performance is \emph{five} seconds.
  \item Only one simultaneous touch is recorded at a time.
  \item Touch gesture data is recorded for replaying the performance.
\end{enumerate}

Such performances require very little effort on the part of users.
While some users may find it difficult to conceive and perform several
minutes of music on a touch-screen device, five seconds is long enough
to express a short idea, but short enough to leave users wanting to
create another recording. It has been argued that five seconds is
enough time to express a sonic object and other salient musical
phenomena.~\cite{Godoy:2010qv}. While the limitation to a single touch
may seem unnecessary on today's multi-touch devices, this stipulation
limits tiny performances to monophony. In order to create more complex
texture or harmony, performers must collaborate, or record multiple
layers themselves.

\begin{table}[]
\caption{An excerpt from a tiny performance data record created in
  MicroJam. These records allow the performance to be recreated from
  touch gestures.}
\label{tab:tiny-perf-example}
\centering
\begin{tabular}{lllll}
\hline
time     & x        & y        & z         & moving \\ \hline
0.007805 & 0.276382 & 0.416080 & 38.640625 & 0      \\
0.065901 & 0.286432 & 0.428141 & 38.640625 & 1      \\
0.074539 & 0.286432 & 0.433166 & 38.640625 & 1      \\
0.090817 & 0.290452 & 0.450251 & 38.640625 & 1      \\
0.107149 & 0.298492 & 0.468342 & 38.640625 & 1      \\
0.124072 & 0.309548 & 0.486432 & 38.640625 & 1      \\ \hline
\end{tabular}
\end{table}

For playback, storage, and transmission to other users, tiny
touch-screen performances are recorded as simple comma-separated
value files of recorded touch gestures. This data format records the
user's performance movements in a compact manner (typical size is
around 5kB), rather than the actual sound or abstract musical values
such as notes and rests. The performance can later be recreated by
sending these same touch event signals to MicroJam's synthesised
instruments. The data format records each touch interaction's time (as
an offset in seconds from the start of the performance), whether the
touch was moving or not, $x$ and $y$ locations, as well as touch pressure
($z$), an example is shown in Table \ref{tab:tiny-perf-example}. The
visual trace of performances is also stored as a PNG image for later
use in the app, although this can also be reconstructed from the touch
data.

Storing gestural control data, rather than audio, or high-level
musical data such as MIDI, allows performances to benefit from updated
synthesis routines in the app, and future enhancements to the visual
interface. As noted in previous work on preserving touchscreen
improvisation~\cite{Martin:2016rm}, this representation encodes
information that might not be available in audio or MIDI recordings.
In Section \ref{studying-tiny-performances} we take advantage of this
information to study tiny performances from a touch-gesture
perspective. 

\subsection{Instruments in MicroJam}\label{instruments}

MicroJam includes eight different instruments (see Table
\ref{tab:instruments}) that map touches in the performance area to
different synthesised sounds. This selection of instruments provides
basic coverage of typical musical roles such as percussion (drums),
bass (fmlead, wub), accompaniment (pad, keys), and lead (chirp,
strings). While far from an exhaustive collection, these instruments
allow exploration of different musical ideas, both through their
different timbres as well as the different touch to sound mappings
used in each one. Descriptions and mapping details for each instrument
is given in Table \ref{tab:instruments}.

The instruments are implemented in Pure Data and make use of the
\texttt{rjlib} library~\cite{Barknecht:2011qe} for some synthesis
routine implementations. They are loaded in the app via
\texttt{libpd}~\cite{Brinkmann:2011fy}. The instruments are controlled
by continuous inputs of $x$, $y$ and $z$ values from the single touch point in
performance area, and each instrument is responsible for its specific
mappings of these values to higher level musical quantities such as
pitch, timbre and control of audio effects. For example, the fmlead
sound, a simple FM synthesiser, maps initial x-values from a tap or
swipe to a bass register pitch. The y-value is mapped to volume. The
initial touch of a swipe triggers a very short note, but continuing a
swipe sustains the sound until the touch is released. For continuous
swipes, the distance of the present touch-point to the initial one ($dx$
and $dy$) is calculated; $dx$ is used to control pitch bend, and $dy$ is
mapped to reverb and delay effects as well as mixing in a copy of the
sound in a lower octave. A similar mapping scheme is used for the other
pitched sounds, see Table \ref{tab:instruments} for precise details.
For the ``drum'' sound, different synthesised drumset sounds (bass
drum, snare drum, hihat, and crash cymbal) are triggered from each
quadrant of the screen and the swipes trigger a roll.

\begin{table}[]
\caption{Descriptions and mapping details of each instrument in
  MicroJam.}
\label{tab:instruments}
\begin{tabular}{lp{4cm}p{1.5cm}p{1.5cm}p{1.5cm}p{2.5cm}}
\hline
instrument & description                                                                                                  & X             & Y                    & dX         & dY                                \\ \hline
chirp      & Basic sine oscillator sound.                                                                                 & pitch         & mix lower octave     & pitch bend &                                   \\
drums      & Drum sounds (bass drum, snare drum, hihat, crash cymbal) triggered from each quadrant. Swipes trigger rolls. & instrument    & instrument           & pitch bend & reverb/delay fx                   \\
fmlead     & Simple two operator FM synthesiser that plays in a bass register.                                            & pitch         & volume               & pitch bend & mix lower octave, reverb/delay fx \\
keys       & Phase modulation keyboard sound.                                                                             & pitch         & volume               & pitch bend & modulation                        \\
pad        & Sawtooth wave pad synth.                                                                                     & pitch         & volume               & pitch bend & tremolo, reverb/delay fx          \\
quack      & Wave packet synth with analogue-like sound.                                                                  & pitch         & timbre               & pitch bend &                                   \\
strings    & Karplus-Strong plucked guitar synthesiser.                                                                   & pitch         & volume               & pitch bend & reverb/delay fx                   \\
wub        & Tremolo ``wub wub'' bass sound.                                                                              & pitch, timbre & tremolo rate, timbre & pitch bend &                                   \\ \hline
\end{tabular}
\end{table}

\section{Studying Tiny Performances}\label{studying-tiny-performances}

In this section we describe an investigation of how users interact
with MicroJam through the lens of the tiny performances that have been
collected in the app. Since early prototypes of MicroJam were
developed, the app has been distributed and demonstrated among
researchers, students, and the music technology community, and these
testers and early users uploaded performances to the app's cloud
database. To date, more than 1600 tiny performances are available in
this database, allowing us to gain insight into the musical potential
of MicroJam and the concept of tiny touchscreen performances.

Our analysis of the tiny performances is made at three levels of
abstraction: whole performances, individual touch events, and gestures
or groups of touch events. At the performance level, we consider
descriptors of each complete 5-second performance and how these vary
by the instrument that was used, we also analyse the visual trace of
the resulting performance. At the touch event level, we consider
individual touch events that may be part of a swipe across the screen,
or single taps. At the gesture level, we consider groups of these
touch events that constitute a single ``note'' or interaction: either
individual taps, or the collection of events that form a swipe. This
lets us examine the different types of movements that users perform.
These levels reveal different aspects of the users' performance
behaviour; while the whole performance level demonstrates their broad
artistic intentions, the gestural and touch levels expose small-scale
interactions.

\subsection{Participants and Data Sources}

Tiny performances for this investigation came from two databases: a
development database which is only accessible from instances of
MicroJam installed on the authors' test devices, and a public database
accessible from beta and published versions of MicroJam. Performances
in the development database were made at testing and demonstration
sessions taking place in our lab, at conferences, and at other events.
Most of the participants in these sessions were university students and
researchers in music technology or computer science.
These participants had a range of music experience from untrained to
professional. The public database was accessed from beta versions of
the app as well as the published app store version. Beta versions were
distributed to interested members of the computer music and technology
community who requested invitations through social media and at
conferences. Since the public release of the app, performances in this
database have been contributed by unknown members of the public.

Unlike most music apps, all performances saved in MicroJam's public
database are openly accessible through the app and through Apple's
CloudKit API. Personal data (e.g., real names, email addresses, IP
addresses) are not included in this data. While the use of public
social media data (e.g., studies of text scraped from Twitter) for
research is commonplace, there remain ethical concerns about whether
users are comfortable with such use~\cite{Fiesler:2018aa}. In our
case, MicroJam's privacy policy highlights our research intentions,
and the number of users from the published version of MicroJam is, so
far, small compared to those who installed the app through private
beta testing.

Within our dataset 39 unique CloudKit users are represented who
created a total of 1626 performances. We can identify performances
created by devices under our control and so know that 431 were created
on the first author's devices and 723 performances are from devices
used for testing and demonstration sessions. The remaining 903
performances are from unknown users. We have chosen to include
performances by the authors as part of an aggregated dataset, although
these could be treated separately as autobiographical
research~\cite{Neustaedter:2012fp}.



\subsection{Performance Level Investigation}

\begin{figure}
  \centering
  \includegraphics[width=1.0\textwidth]{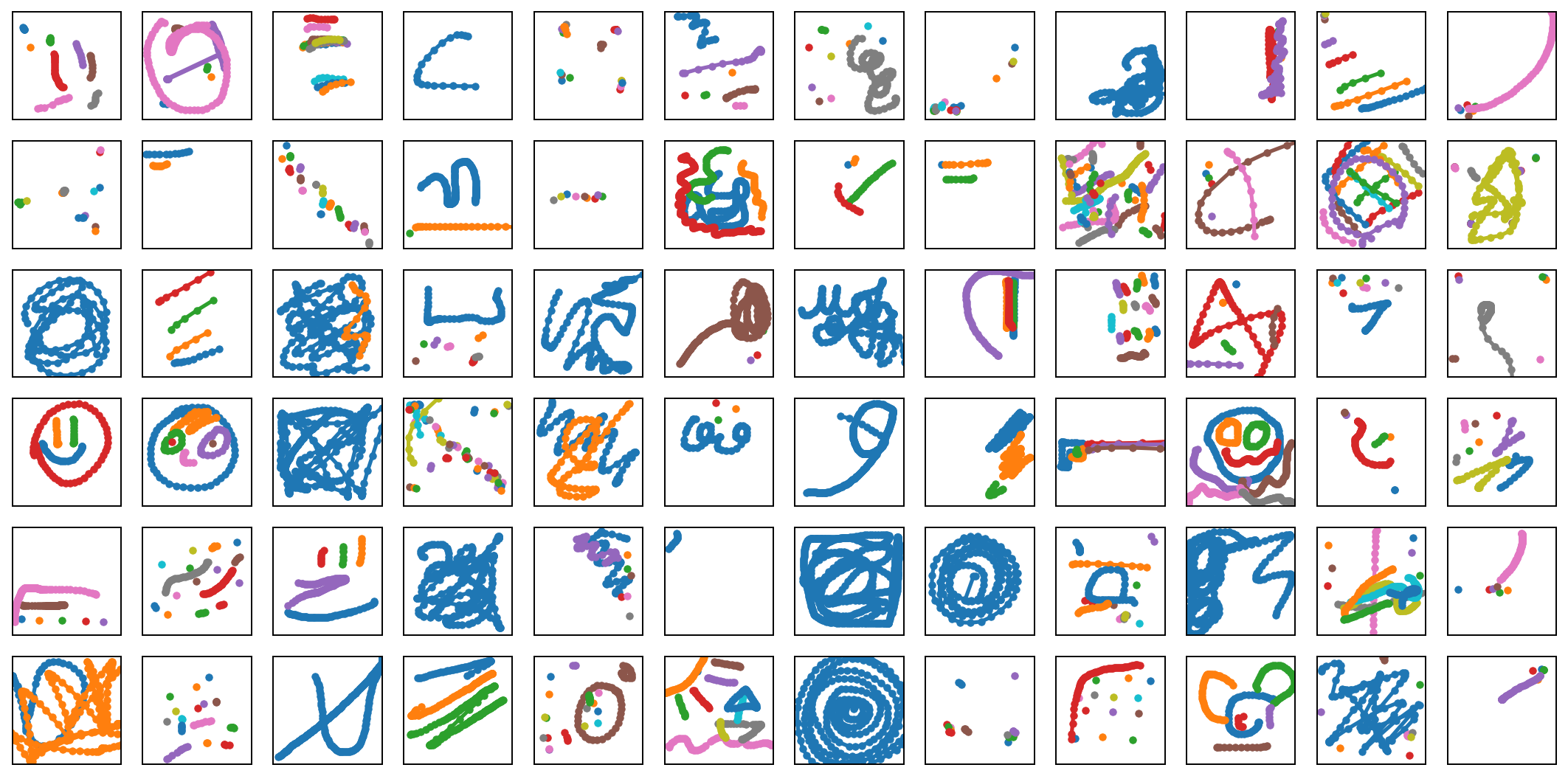}
  \caption{The visual trace of a selection of MicroJam performances
    with each swipe coloured differently. Some different performance
    styles are visible, for example: taps, single long swipes,
    repeated strokes, and drawing an image. Some of the interactions
    appear to be predominantly visual, while others are exploratory
    scribbles with little visual coherence.}
  \label{fig:tiny-performances}
\end{figure}

\begin{figure}
  \centering
  \includegraphics[width=0.5\textwidth]{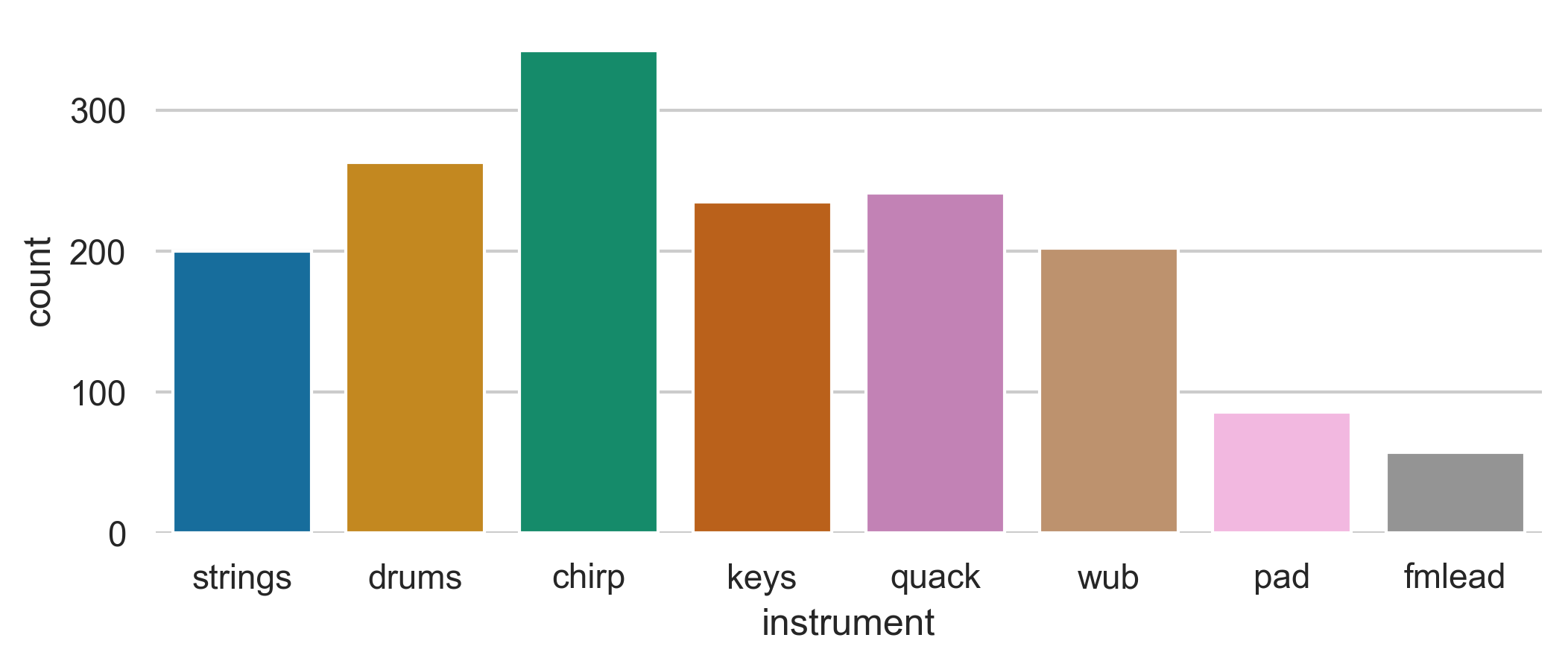}
  \caption{Number of performances with each instrument in our
    dataset. The default instrument, ``chirp'' is most popular
    followed by drums. ``fmlead'' and ``pad'' are less popular,
    potentially due to a less appealing synthesis mapping and sound.}
  \label{fig:counts-by-performance}
\end{figure}

The visualisations of a subset of tiny performances are reproduced in
Figure \ref{fig:tiny-performances}. This figure shows the variety of
touch-interaction styles that have been generated by performers. Many
of the interactions are abstract, resembling scribbles that show the
user experimenting with the synthesis mapping of the jamming
interface. Some performances contain patterns where performers have
repeated rhythmic motions in different parts of the touch area. A
number of the performances are recognisable images: figures, faces,
and words. We can characterise the visual appearance of performances
under the following broad styles: taps, swipes, long swipe, mixture,
image, and text. Tap and swipe performances focus on these fundamental
touch gestures, while long swipe performances seem to consist of only
one swipe, and mixture performance include multiple of these styles.
Image and text performances appear to focus on the visual meaning of
the finished trace, likely with less emphasis given to the resulting
sound.

Of the 1626 performances in the dataset, 479, or 29\%, are replies. As
replies of replies are possible in MicroJam, it is interesting to see
how long potential chains of replies are in terms of number of
performance layers. Of the 479, 361 have just two layers. While there
are 83 performances with 3 layers, there are few with 6 and 7 layers
(3 each). This data suggests that while users have made some use of
the reply function, they have only rarely explored the potential for
complex layered performances. Further work in developing the browsing
screen could help encourage users to create longer performances; for
instance, the number of layers in a performance could be shown, with
complex performances highlighted while browsing.

While performances with all instruments are represented in the
dataset, the most popular instrument is chirp (the default choice)
with 342 performances. Figure \ref{fig:counts-by-performance} shows
that the next five most popular instruments (drums, keys, quack,
strings, wub) all have around 200 performances, but that fmlead and
pad each have less that 100. It is not immediately clear why these two
instruments are less popular, but it may be that their sound is less
distinctive that the other instruments, particularly through a mobile
device's small speaker.



\subsection{Analysis by Touch Event}

The dataset includes 249,870 touch events. As set out in the tiny
performance definition in Section \ref{tiny-performance}, a touch
event can either be a \emph{touch-down} (when the user's finger hits the
touchscreen), or a \emph{touch-moved} (when the user's finger has moved
without leaving the touchscreen). These are distinguished by a
binary value, ``moving'', in the dataset. Only 13,480 touch events in
the dataset are non-moving compared to 236,390 moving touches, this is
because whenever a user swipes during a performance a large number of moving
touches are generated, while only one touch-down occurs. 

\begin{figure}
\centering
\begin{subfigure}{.5\textwidth}
  \centering
  \includegraphics[width=0.99\linewidth]{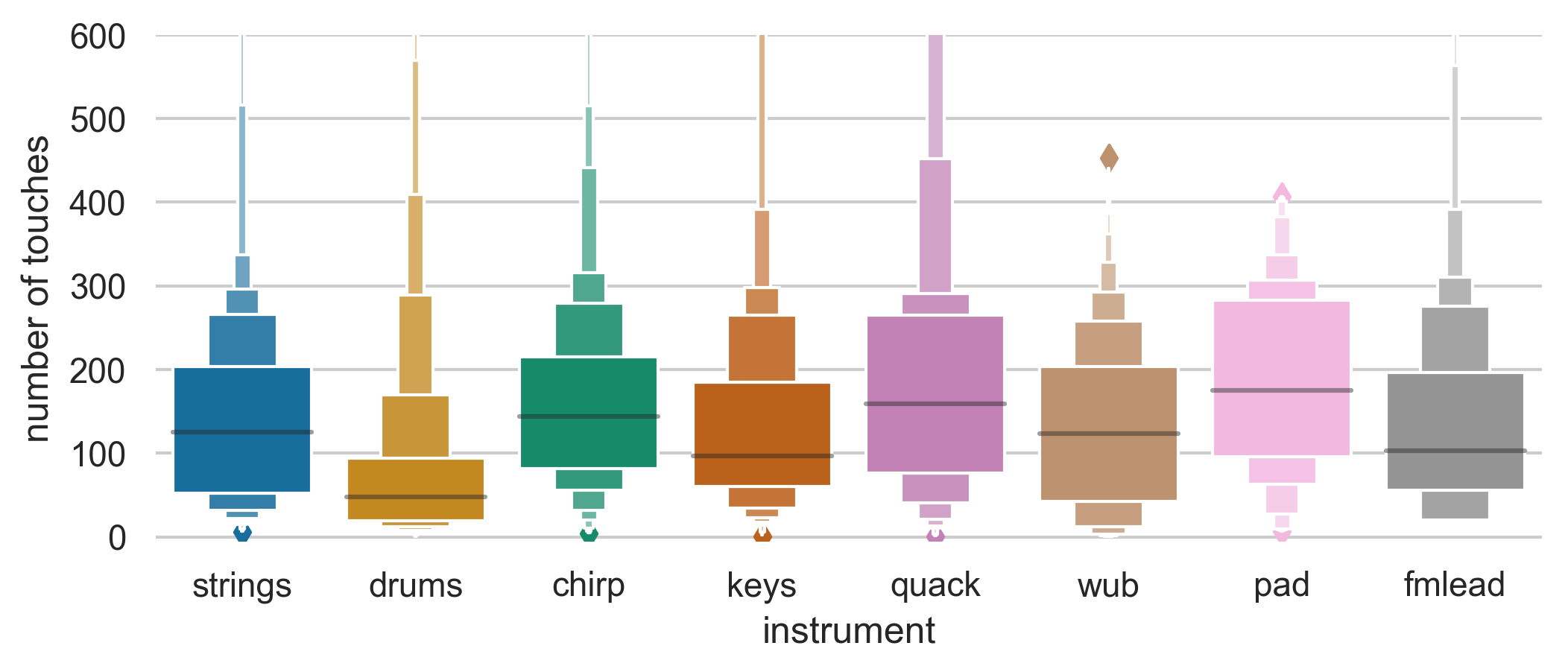}
  \caption{Distribution of the total count of touchpoints in each
    performance for each instrument.}
  \label{fig:count-touches}
\end{subfigure}%
\begin{subfigure}{.5\textwidth}
  \centering
  \includegraphics[width=0.99\linewidth]{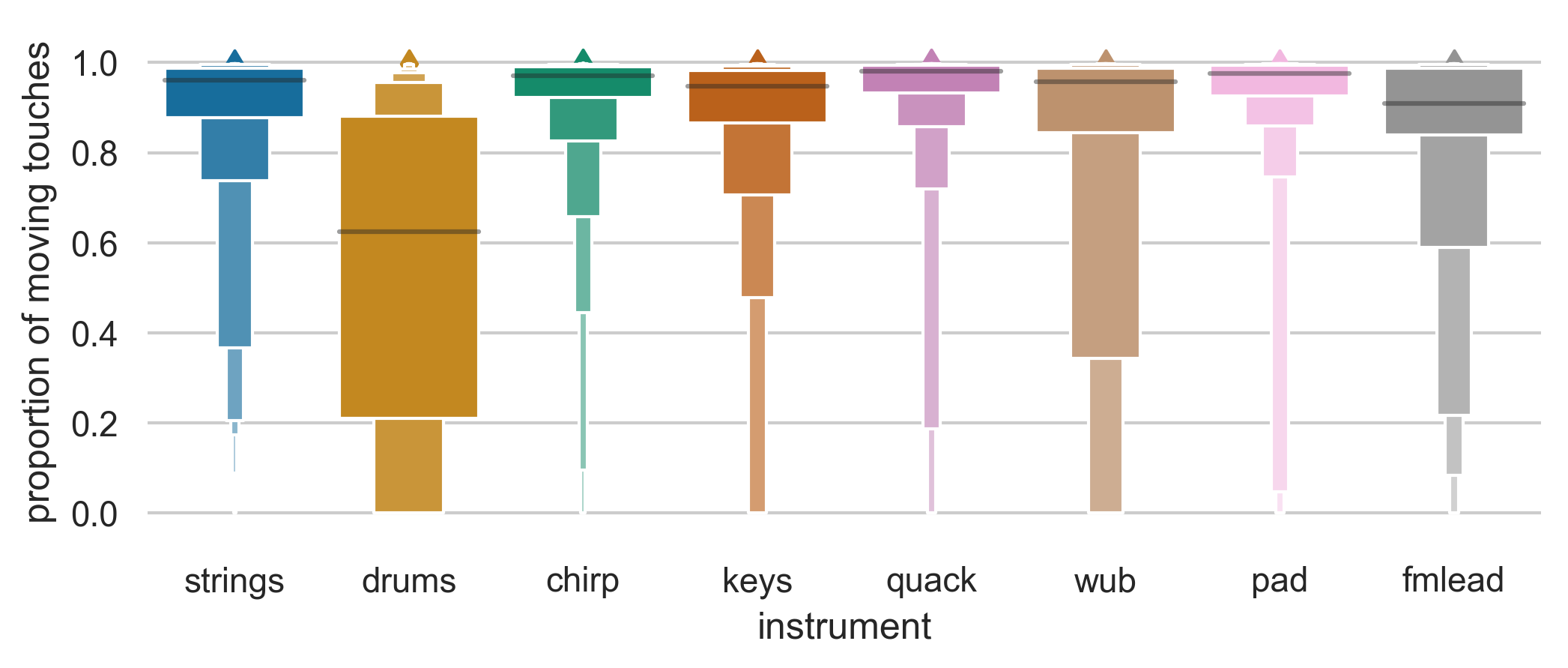}
  \caption{Distribution of moving vs. non-moving touchpoints for each instrument.}
  \label{fig:moving-proportion}
\end{subfigure}
\caption{Distributions of touchpoint properties divided by instrument
  shown as letter-value plots~\cite{Hofmann:2017aa}. Each
  instrument had a similar number of touchpoints per performance with
  somewhat fewer for drums. All instruments except drums are dominated
  by moving touches.}
\label{fig:pointwise-properties}
\end{figure}

Figure \ref{fig:count-touches} shows the distribution of the number of
touch events recorded per performance divided by instrument used. This
shows that for all instruments except drums performances had a median
of between 100 and 200 touches. The median number of touch events for
drums was 48, much lower than the other instruments. This is explained by Figure
\ref{fig:moving-proportion} which shows the proportion of touch events
that were moving in each performance. For drums, many more touch
events were taps, rather than swipes, which resulted in fewer touch
events for a given performance.

\begin{figure}
  \centering
  \includegraphics[width=0.5\textwidth]{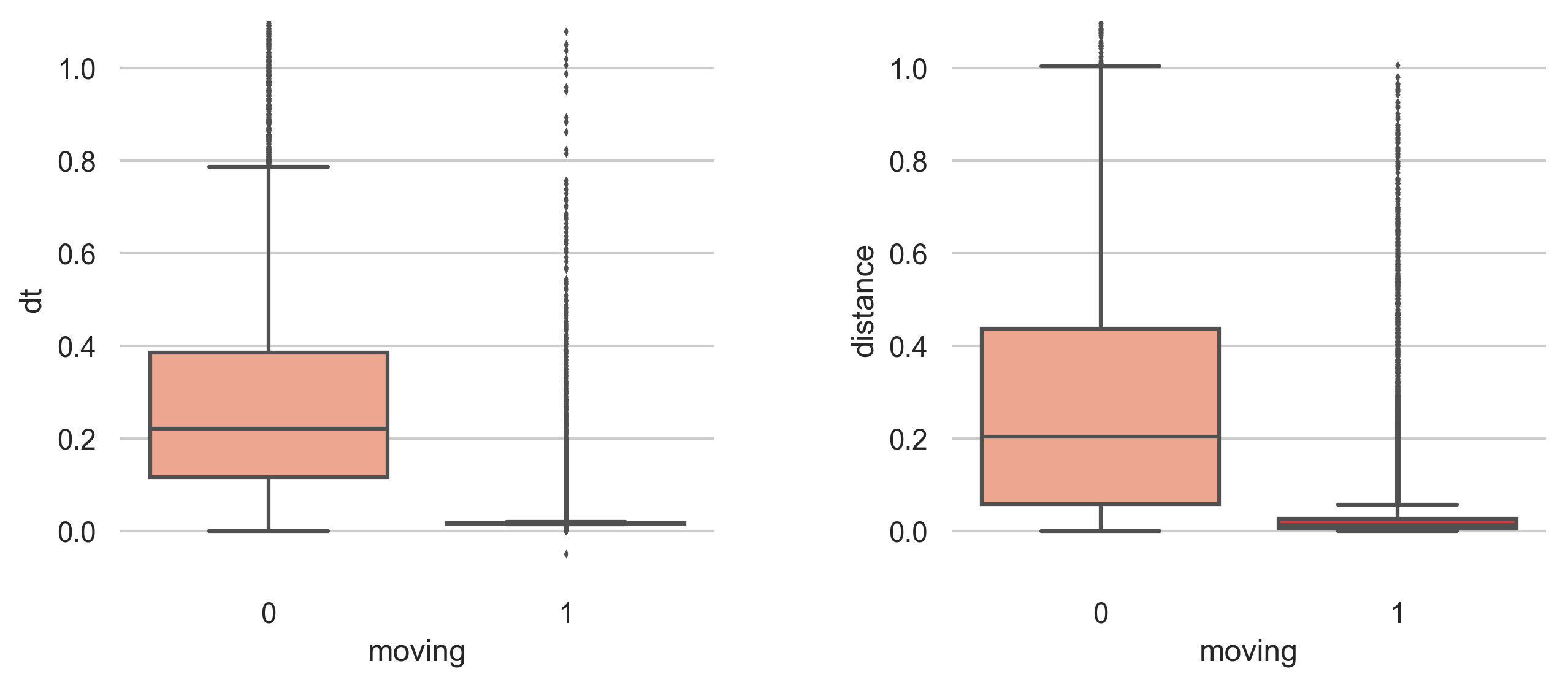}
  \caption{Distributions of touch event dt values (left) and screen
    distance (right) divided by whether the touch was moving or not.}
  \label{fig:pointwise-dt-dxy}
\end{figure}

Two interesting statistics are the time differences between
consecutive touch events ($dt$), and the onscreen distance between
them measured as a proportion of the performance area. Distributions
of these statistics are shown in Figure \ref{fig:pointwise-dt-dxy}. As
expected, values of $dt$ and distance for moving touches tend to be
small, although there are many outliers. The median $dt$ between
moving touches is only 0.017s compared with 0.221s for non-moving
touches. The interquartile range of $dt$ for non-moving touches is
0.116s---0.385s, this gives an indication that performers tend not to
leave much time in between interactions and the resulting tiny
performances would not have much temporal ``space''. Similarly, we can
observe that non-moving touches tend to have moved within a relatively
small proportion of the performance area (median = $0.203$,
interquartile range = $[0.058,0.436]$). This can be observed in some
of the example performances in Figure \ref{fig:tiny-performances} where
a user has tapped repeatedly in the same part of the performance area.

\begin{figure}
  \centering
  \includegraphics[width=0.4\textwidth]{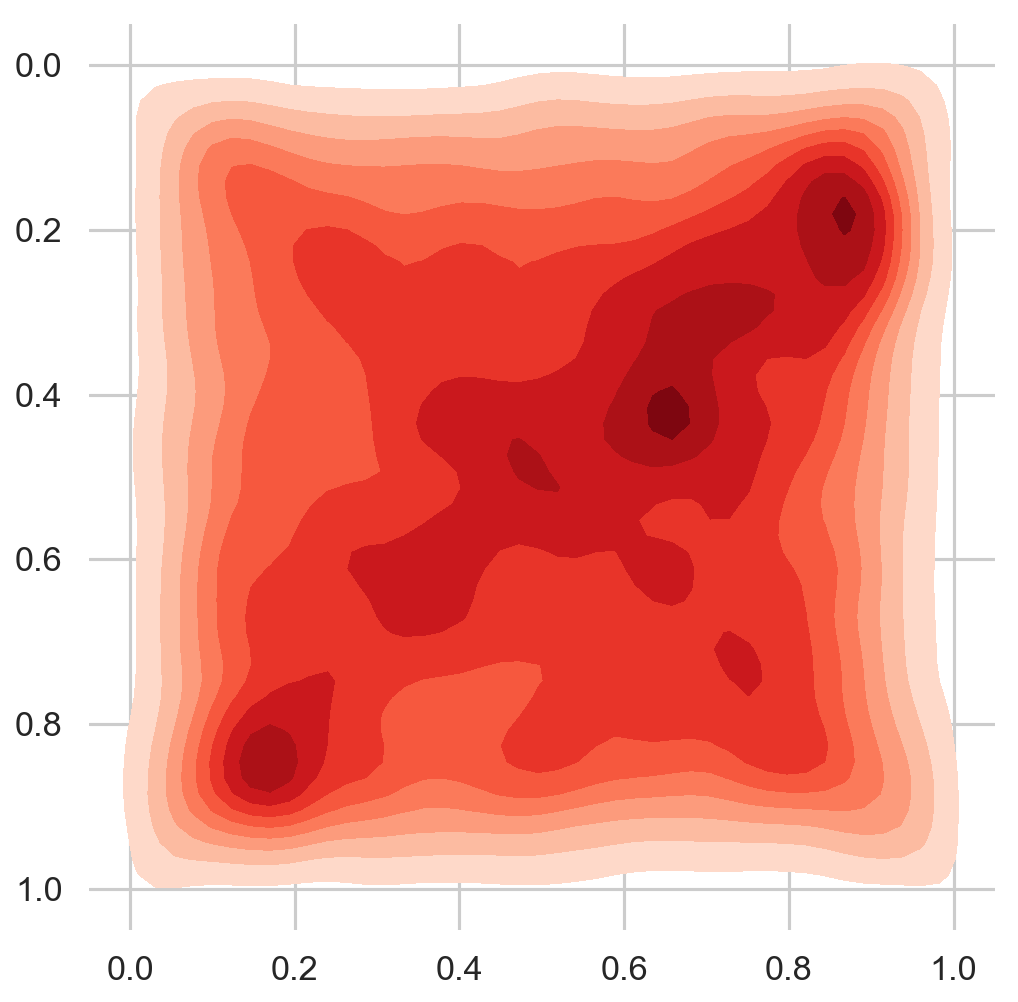}
  \caption{The distribution of touch event locations across the touchscreen
    performance area (bivariate kernel density estimation) where darker
    colours indicate higher touch event density. Touches broadly cover
    the whole performance area, but are more prominent on the
    diagonals and particularly in the upper right corner.}
  \label{fig:pointwise-kde}
\end{figure}

Figure \ref{fig:pointwise-kde} shows the distribution of touch events
across the touchscreen performance area (obtained with a bivariate
kernel density estimate). This allows us to investigate where users
have most commonly interacted with the performance area to play
sounds. We can observe that touches cover the whole performance area;
however, more touches occur on the diagonals and in the upper right
quadrant. Relatively few touches extend all the way to the edges.

These analyses suggest that users tend not to explore the potential of
space in their performances, both in terms of time and the touchscreen
area. Given the time limitation of five seconds, it is understandable
that users would prefer to squeeze in as much activity as possible.
However, a small number of interactions could also be effective by
allowing pauses that add structure to the
performance~\cite{Sutton:2002aa}. Taken together, these results could
inform future synthesis mappings in the app. For instance, a mapping could produce
unexpected or interesting sounds if the next touch event is far
away in space or time. Given that users tend not to use the edge of
the performance area, these areas could be mapped to more extreme
sounds (e.g., with distortion or delay effects). Similar mappings have
been explored in instruments such as
``Crackle''~\cite{2011-Crackle-Reus}.

\subsection{Analysis by Gesture}

In this section we analyse performances from the perspective of tap
and swipe gestures, the two fundamental touchscreen interactions
available in MicroJam's interface. A swipe is a sequence of multiple
touch events that can be defined as a touch-down followed by a
non-zero number of touch-moved events. The dataset contains 7090 such
swipes, compared with 6390 true taps where a touch down was not
followed by a touch moved event. These swipes represent one of the
more important phenomena in tiny-performances as they represent the
actions formed by the majority of touch-points. We extracted swipes
from each performance in the dataset by dividing them by touch down
events and discarding all divisions with only a touch-down. 155 swipes
that were longer than 5 seconds were excluded. We were then able to
perform analyses that characterise swiping behaviour seen in our
dataset.

\begin{table}[]
  \caption{Descriptive statistics of swipes in the dataset. Most
    swipes are short in time (75\% are less than 0.54s) and distance
    (75\% are trace less than 0.60 of the performance area). While swipes that
    cover a long distance and a long period of time dominate the
    visual traces of performances, they are in the
    minority in terms of user behaviour.}
\label{tab:swipe-data}
\centering
\begin{tabular}{llllll}
\hline
statistic  & length (events)     & time (s)  & mean velocity& distance  & max velocity\\
\hline
mean       & 27.56               & 0.4976    & 1.7774       & 0.6006    & 5.8506\\
std        & 48.04               & 0.8768    & 4.9702       & 1.4933    & 27.7702\\
min        & 2                   & 0.0012    & 0.0          & 0.0       & 0.0\\
25\%       & 4                   & 0.0605    & 0.3879       & 0.0310    & 0.9189\\
50\%       & 7                   & 0.1284    & 0.8710       & 0.1582    & 1.9634\\
75\%       & 29                  & 0.5390    & 1.5843       & 0.5977    & 3.8192\\
max        & 399                 & 4.9998    & 102.2893     & 49.8778   & 1600.4297\\
\hline
\end{tabular}
\end{table}

Table \ref{tab:swipe-data} shows descriptive statistics on the 6935
valid swipes. The majority were short in time and on-screen distance
(measured in proportions of the performance width covered). 75\% of
swipes were shorter than 0.54s in duration and the median swipe time
was only 0.13s. The median distance was 0.16 area widths and 75\% of
swipes traced less than 0.6 of the area width. These results seem
inconsistent with the appearance of the visual traces shown in Figure
\ref{fig:tiny-performances} where the images appear to be dominated by
long continuous swipes. The dataset does contain long swipes, up to
the whole 5s in time and almost 50 area widths. These long swipes are
visually dominant, especially given that they can cover up some
smaller interactions, but they are outnumbered by the shorter swipes
that make up the vast majority of gestures.

\begin{figure}
\centering
\begin{subfigure}{.5\textwidth}
  \centering
    \includegraphics[width=0.95\linewidth]{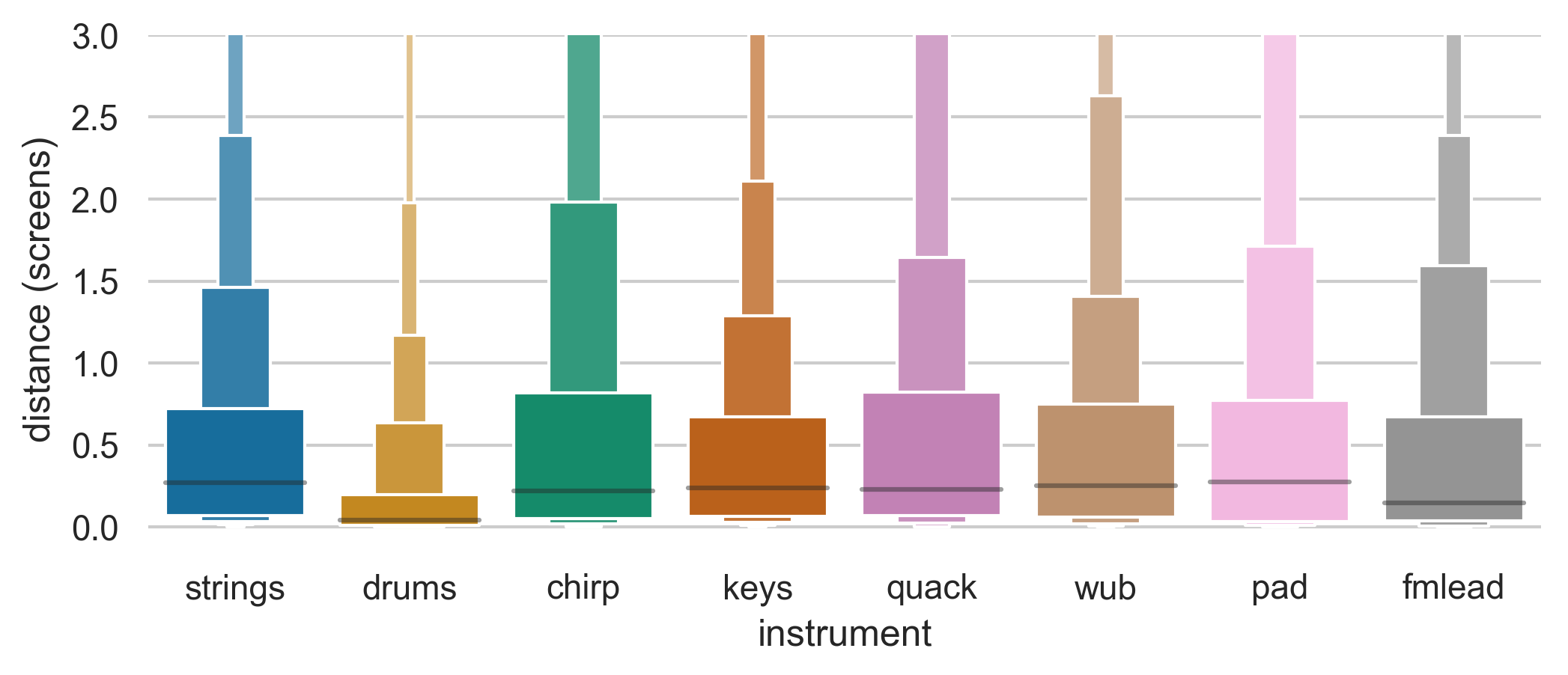}
  \caption{Distance travelled over each swipe.}
  \label{fig:swipe-distance}
\end{subfigure}%
\begin{subfigure}{.5\textwidth}
  \centering
  \includegraphics[width=0.95\linewidth]{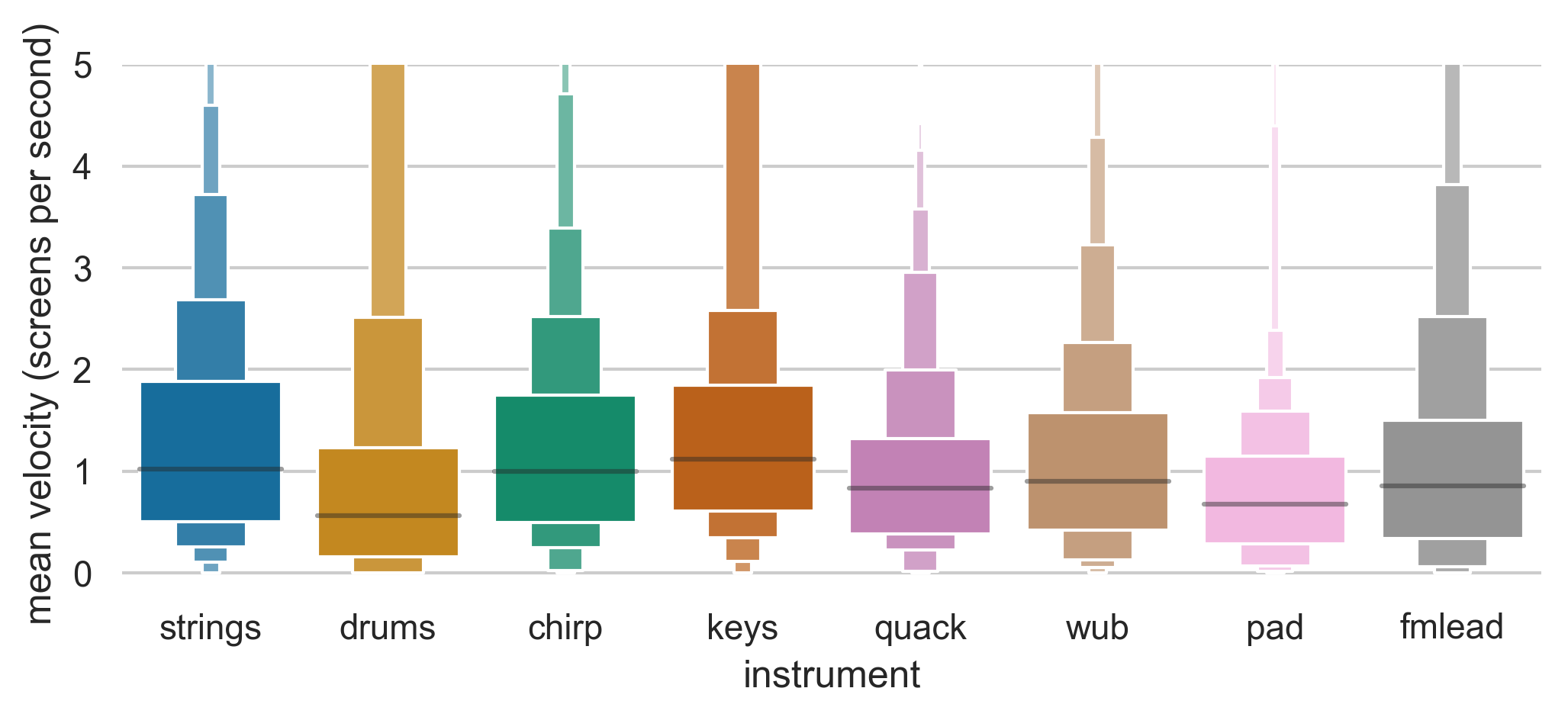}
  \caption{Mean velocity over each swipe.}
  \label{fig:swipe-mean-vel}
\end{subfigure}
\vspace{5mm}
\begin{subfigure}{.5\textwidth}
  \centering
  \includegraphics[width=0.95\linewidth]{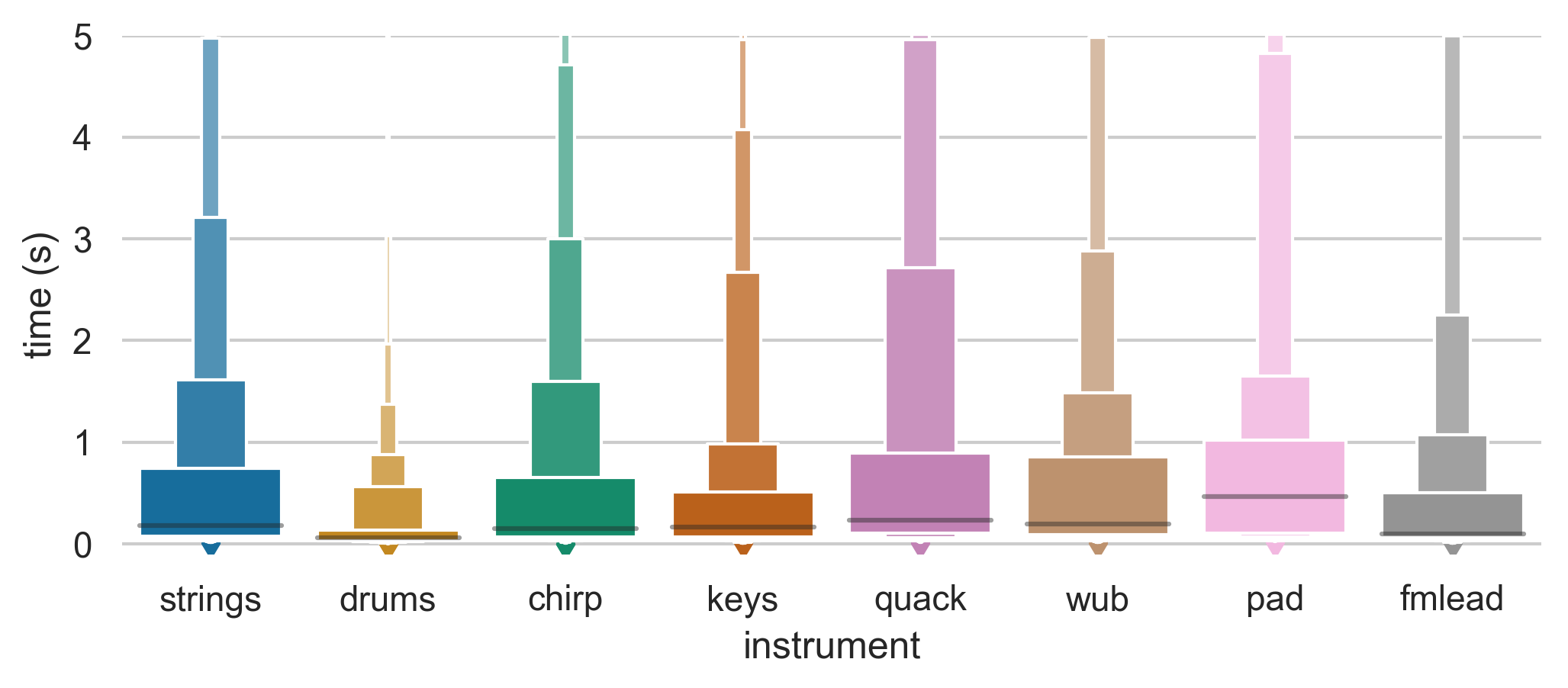}
  \caption{Time taken for each swipe.}
  \label{fig:swipe-time}
\end{subfigure}
  \caption{The distribution of swipe statistics for different
    instruments shown as letter-value plots~\cite{Hofmann:2017aa}.}
  \label{fig:swipe-statistics}
\end{figure}

In Figure \ref{fig:swipe-statistics}, we show the distributions of
time, distance, and mean velocity for each instrument in MicroJam.
One-way ANOVA tests on each measurement confirm that there are
significant effect due to instrument ($p < 0.001$). In particular,
drum performances have much shorter swipes than any of the other
instruments in terms of both distance traced and time. This could be
due to many attempted ``taps'' that were actually short swipes with
just a few touch-points. Swipes using the pad sound seem to be long in
terms of time, but with a lower mean velocity, indicating the use of
this instrument to play long notes with slower variation in pitch and
timbre, perhaps by subtle movement in a small area.

\begin{figure}
  \centering
  \includegraphics[width=0.23\textwidth]{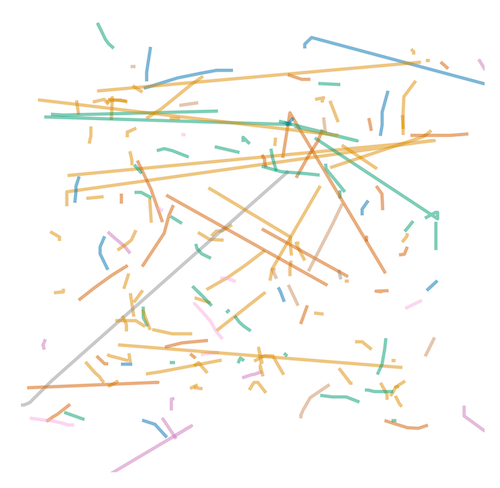}
  \includegraphics[width=0.23\textwidth]{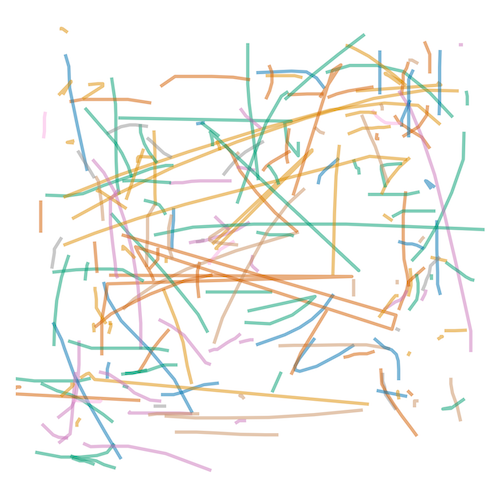}
  \includegraphics[width=0.23\textwidth]{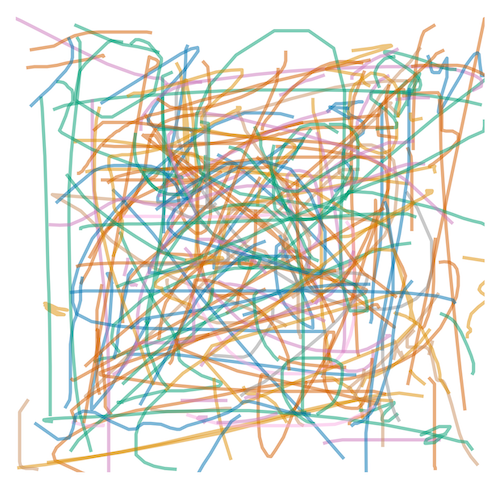}
  \includegraphics[width=0.23\textwidth]{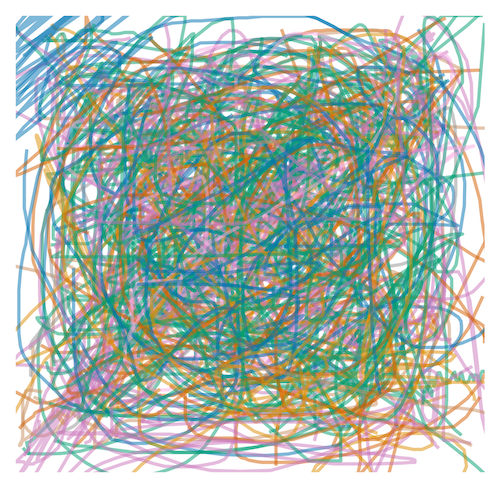}
  \caption{Random selections of 200 swipes from each quartile of
    swipe-length (in time) represented in the dataset. Each swipe is
    coloured by the instrument used to generate it.}
  \label{fig:swipe-selections}
\end{figure}

To gain an intuitive idea of how these swipes looked, we have
visualised selections of swipes of different time-length, these are
shown in Figure \ref{fig:swipe-selections}. First, the quartiles for
the time dimension were calculated (see Table \ref{tab:swipe-data}),
then 200 swipes were sampled from each quartile randomly. The quickest
25\% of swipes are almost always straight lines with many spanning
quite far across the screen. Some of these very long swipes may
indicate bugs in the interface code that has failed to reject multiple
touch points on the screen, instead registering them as a single
swipe. The two quartiles around the median length show much variation
in expression. Swipes shorter than the median
rarely have more than a subtle curve, while those above the median
show curves that could have an expressive effect on the pitch and
timbre of the resulting sounds. In the upper quartile of length,
swipes can cover the whole performance area or trace complex patterns.
These swipes could represent longer notes, parts of drawings, or whole
5-second performances.

\begin{figure}
  \centering
  \includegraphics[width=0.45\textwidth]{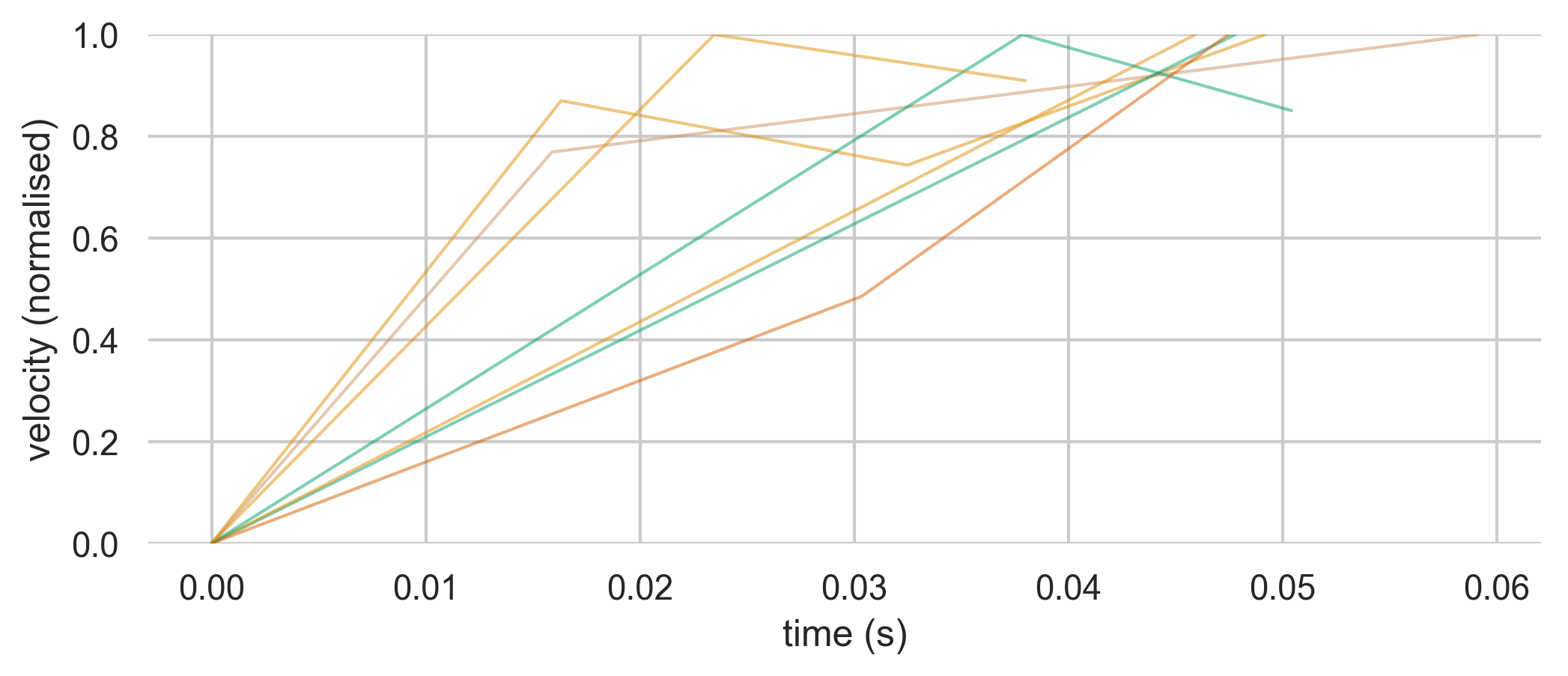}
  \includegraphics[width=0.45\textwidth]{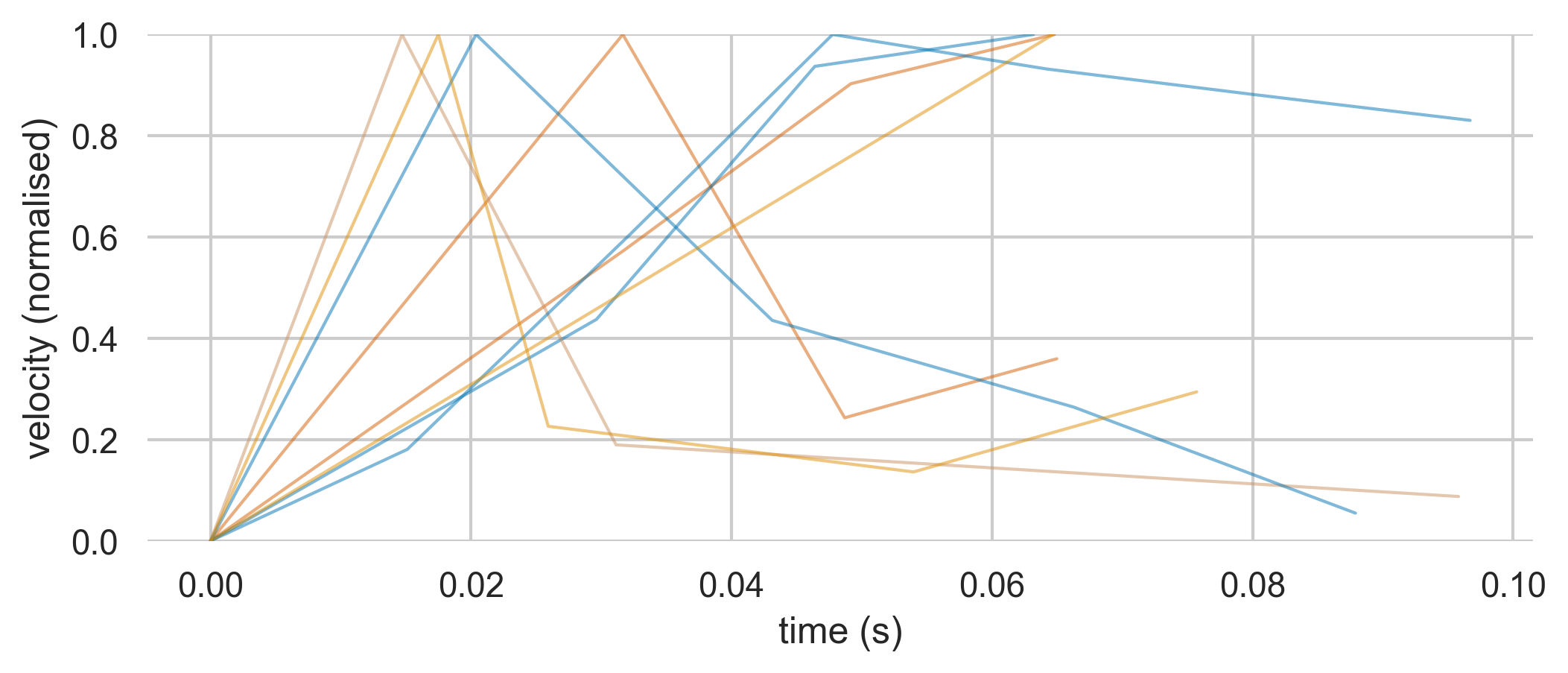}
  \includegraphics[width=0.45\textwidth]{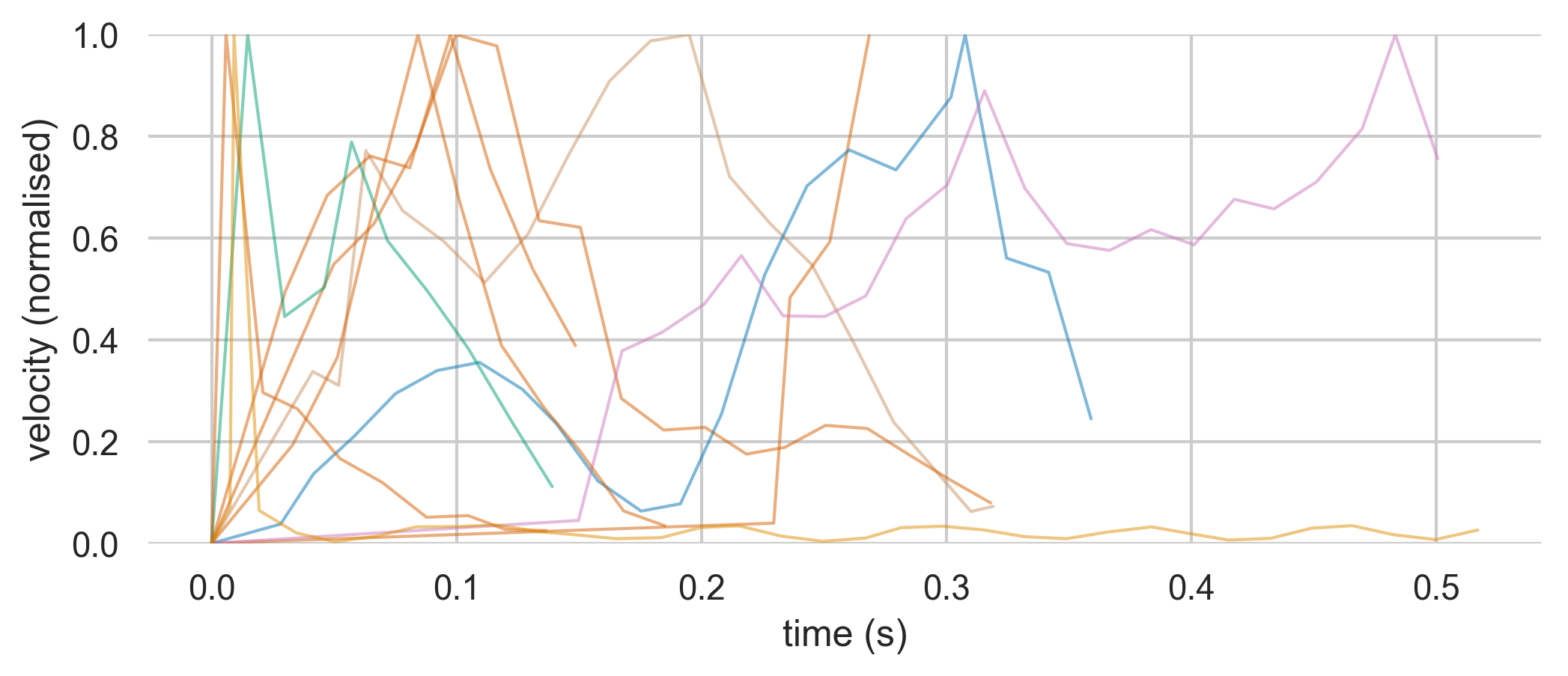}
  \includegraphics[width=0.45\textwidth]{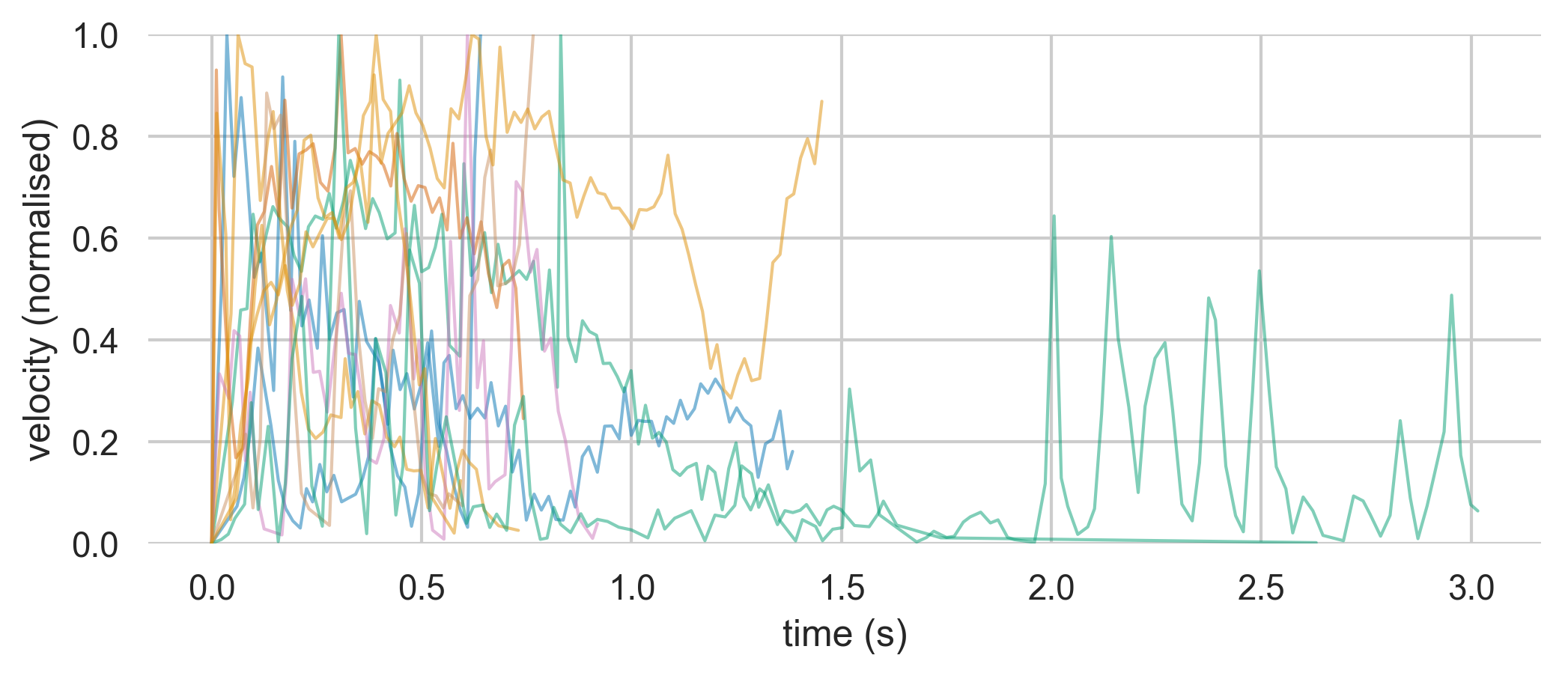}
  \caption{The normalised velocity curves of random selections of 20
    swipes from each time quartile represented in the dataset. Each
    curve is coloured by the instrument used to create it. The
    shortest swipes tend to be strictly increasing in velocity while
    the longer two quartiles show some temporal activity such as
    pulsing velocity that could indicate tracing back and forth
    across the screen.}
  \label{fig:swipe-velocities}
\end{figure}

The 2D traces of performances in figures \ref{fig:tiny-performances}
and \ref{fig:swipe-selections} do not show the velocity of swipes---a
quantity with much expressive potential. In Figure
\ref{fig:swipe-velocities}, we visualise the normalised velocity
curves for selections of swipes of different lengths. The shortest
swipes have only 2 or 3 touch points and the velocity tends to only
increase, indicate a quick flicking movement. The next quartile shows
a more expressive curve, with a quick rise and slower release as the
touch point stops moving before the end of the swipe. The third
quartile shows the possibility for multiple peaks and valleys in the
velocity, perhaps indicating changes in direction of the moving touch
point. Again, only the fourth quartile shows extensive expressive
behaviour such as repeating peaks in velocity that could indicate a
rhythmic movement pattern over a longer swipe.

The analysis of swipes in MicroJam has left us some important
insights. Most importantly, the majority of swipes are short---three
quarters are less than 0.54 seconds---however, the longest swipes have
more scope for expressive behaviour. This has implications for future
instrument design that should allow more detailed expression with very
short interactions, e.g., by slowing down the effect a swipe has on
audio effects or timbral changes to make them more noticeable.
Alternatively, long swipes could be encouraged. It could be if the
visual impact of long swipes was reduced by fading them out over time,
users might feel more confident to explore them more frequently.
Further refinements to instruments such as ``drums'' that reward
longer swipes with interesting sounds (e.g., pitch bends on toms or
cymbals) could also encourage such behaviour.

\subsection{Discussion}


The results above allow us to characterise how users perform within
the constraints of MicroJam's tiny performance idiom. We know that
they perform in different broad styles including abstract gestures as
well as images and text. While visually meaningful performances draw
the eye, the dataset is dominated by very short swipes and taps which
are typical of the more abstract performance styles. It is
questionable whether image and text explorations lead to rewarding
musical experiments, and it may be more appropriate to focus on
improving the expressive potential of other performance styles. As for
the social aspects of MicroJam, while the reply function has certainly
been used in the dataset, few multi-layered performances are present
which limits the conclusions we can draw. Future revisions of MicroJam
could emphasise replying and collaboration rather than just
performance creation.

The results of our analysis lead us to make the following design
recommendations that could be explored in future versions of MicroJam
and other similar DMIs:

\begin{enumerate}
\item Multi-layered performances should be encouraged and celebrated
  within the app. At present, few performances have more than two
  layers. To encourage users to create more complex performances,
  these performances could be highlighted in the world feed, and
  opportunities to reply presented more actively.
\item The edge of the performance area should sound edgy. The edges
  are rarely used by performers; mappings could be altered to use
  these spaces to create more exotic or experimental sounds that
  reward the user's exploration.
\item Short swipes should be a focus for improvement to synthesis
  mappings. These interactions are the most common gesture, and should
  be emphasised more in the sound design for MicroJam.
\item The visual impact of long swipes should be reduced. These
  interactions are rare, yet dominate the visual trace of
  performances. Perhaps fading out long swipes would allow multiple of
  them to be used in performances without overwhelming the touch area.
\end{enumerate}

So far, MicroJam has mainly been used in test and demo environments,
and few users have shared large numbers of performances. As a result,
the performances analysed in this dataset are generally by
inexperienced users. We would expect, however, that as for other
instruments, MicroJam users would improve with practice, and develop
new styles. Future work could seek to identify changes in tiny
performance style over time.

One aspect of analysis that has not been mentioned is modelling and
generation of tiny performances with machine-learning algorithms.
Previous research has already discussed a mixture density recurrent
neural network model for generating and responding to tiny
performances~\cite{Martin:2018ag}. This ``RoboJam'' system is
available in the app to provide an automatic reply to a performance on
demand and generates the same control data format as the tiny
performances. In future work we could explore the potential for developing
tailored models of individual users' styles which could even provide
control over the kind of gestures (for instance, short swipes or taps)
that are provided in an automatic response. These models could also be
used to demonstrate effective use of the gestures highlighted in the
above recommendations as part of a performance training feature.

\section{Conclusions and Future Work}\label{conclusions}

In this paper we have presented the design for MicroJam, a social
mobile music app for creating touchscreen performances and defined the
tiny performance format. We have also investigated
this app through a data-driven analysis of more than 1600
performances. This investigation has revealed how users perform in the
tiny touchscreen idiom and allowed us to make recommendations for
revisions that could better align the app's capacity for musical
expression with user behaviour. MicroJam is an example of a social app
centred on musical creation rather than written and visual media. We
have argued that such apps could take advantage of the ubiquity of
mobile devices by allowing users to collaborate asynchronously and in
different locations, and shown that these modes of interaction are
relatively unexplored compared to more conventional ensemble
performances.

MicroJam represents a new approach to asynchronous musical
collaboration with the focus on time-limited tiny performances. Taking
inspiration from the constrained contributions that typify social
media apps, MicroJam limits users to five-second touchscreen
performances, but affords them extensive opportunities to browse,
playback, and collaborate through responses. MicroJam's tiny
performance format includes a complete listing of the touch
interactions and so allows performances to be easily distributed,
visualised, and studied.


Our novel data-driven investigation examined 1626 performances
consisting of 249,870 touch events. These were analysed at the levels
of individual touch events, grouped touch gestures, and whole
performances. The investigation revealed the variety of
styles used in performances but that fewer performances than desired
were replies. Examining touch points showed that the edges of the
performance area was not used as much as the centre and main
diagonals, and that moving touches, rather than taps, dominated the
dataset. Grouping touches into swipes showed that while long swipes
are more visually apparent, the vast majority of swipe gestures are
actually short.

We have distilled the findings of our investigation into
design recommendations for enhancing the instrument mappings and
visualisations in MicroJam. These could encourage users to be more
expressive, and future work could explore how these enhancements
affect tiny performances. Given the social goal of MicroJam, the most
important measure could be to encourage users to interact through
musical collaboration and to generate more complex sequences of
replies. MicroJam could potentially host very large collaborations
between users and performances with multiple threads of replies.
Automatic traversal of such structures could constitute a kind of
generative composition with users' original musical material.

The analysis in this article has suggested that even a simple and
constrained touchscreen interface can lead to a variety of styles and
unexpected musical interactions. While constraining the length of
performances may have increased the number recorded, and made it
easier to collaborate using musical replies, it could have curbed
users' gestural exploration with the touch screen. Implementing the
design recommendations may encourage more expressive performances in
MicroJam users, without increasing the effort required to generate
tiny touchscreen performances. For music making, as opposed to
appreciation, to be widely adopted as part of everyday social media
interactions, this balance between constraint and expression will need
to be further examined and addressed. We posit that a data-driven
approach to mobile music performance, examining musical data generated
by users, can be used to further examine this balance in MicroJam and
other systems for mobile musical creativity.

\subsubsection*{Funding}
This work was partially supported by The Research Council of Norway
through the Engineering Predictability with Embodied Cognition (EPEC)
project, under grant agreement 240862, and the Centres of Excellence
scheme, project number 262762.

\subsubsection*{Acknowledgements}
We wish to thank participants in our study as well as beta testers and
others who tried MicroJam. We also thank Henrik Brustad and Benedikte
Wallace who worked on MicroJam as research assistants.

\subsubsection*{Conflict of Interest}
The authors declare no conflict of interest.

\bibliographystyle{ACM-Reference-Format}
\bibliography{references}

\end{document}